\documentclass[%
 reprint,
 superscriptaddress,
 amsmath,amssymb,
 prb,
]{revtex4-2}

\usepackage{graphicx}
\usepackage{dcolumn}
\usepackage{bm}
\usepackage{graphicx}
\usepackage{color}
\usepackage[colorlinks, linkcolor=blue, anchorcolor=blue, citecolor=blue, urlcolor=blue]{hyperref}
\usepackage{amsmath}
\usepackage{multirow}
\usepackage{tabularx}
\usepackage{longtable}
\usepackage{rotating}

\UseRawInputEncoding

\emergencystretch=\maxdimen
\newcommand{\KMSO}{K$_2$Mn(SeO$_3$)$_2$}

\begin{document}


\title{Directional selection of field-induced phases by weak anisotropy in triangular-lattice \KMSO}


\author{Bin Wang}
\affiliation{Guangdong Provincial Key Laboratory of Magnetoelectric Physics and Devices, Center for Neutron Science and Technology, School of Physics, Sun Yat-Sen University, Guangzhou 510275, China}
\affiliation{School of Physical Sciences, Great Bay University, Dongguan 523808, China}
\affiliation{Songshan Lake Materials Laboratory, Dongguan 523808, China}
\author{Yantao Cao}
\affiliation{Institute of Physics, Chinese Academy of Sciences, Beijing 100190, China}
\affiliation{Songshan Lake Materials Laboratory, Dongguan 523808, China}
\author{Andi Liu}
\affiliation{School of Physics and Wuhan National High Magnetic Field Center,
Huazhong University of Science and Technology, Wuhan 430074, China}
\affiliation{Songshan Lake Materials Laboratory, Dongguan 523808, China}
\author{Guoliang Wu}
\affiliation{Institute of Theoretical Physics, Chinese Academy of Sciences, Beijing 100190, China}
\author{Jin Zhou}
\affiliation{School of Physics and Wuhan National High Magnetic Field Center,
Huazhong University of Science and Technology, Wuhan 430074, China}
\author{Xiaobai Ma}
\affiliation{Neutron Scattering Laboratory, Department of Nuclear Physics, China Institute of Atomic Energy, Beijing 102413, P. R. China}
\author{Wenyun Yang}
\affiliation{State key Laboratory of Artificial Microstructure and Mesoscopic Physics£¬School of Physics, Peking University, Beijing 100871, P. R. China}
\author{Takashi Ohhara}
\affiliation{J-PARC Center, Japan Atomic Energy Agency, Tokai, Ibaraki 319-1195, Japan}
\author{Akiko Nakao}
\affiliation{Neutron Science and Technology Center, CROSS, Tokai, Ibaraki 319-1106, Japan}
\author{Koji Munakata}
\affiliation{Neutron Science and Technology Center, CROSS, Tokai, Ibaraki 319-1106, Japan}
\author{Bing Shen}
\affiliation{Guangdong Provincial Key Laboratory of Magnetoelectric Physics and Devices, Center for Neutron Science and Technology, School of Physics, Sun Yat-Sen University, Guangzhou 510275, China}
\author{Zhendong Fu}
\affiliation{Songshan Lake Materials Laboratory, Dongguan 523808, China}
\author{Zhaoming Tian}
\affiliation{School of Physics and Wuhan National High Magnetic Field Center,
Huazhong University of Science and Technology, Wuhan 430074, China}
\author{Qian Tao}
\affiliation{School of Physics, Zhejiang University, Hangzhou 310027, China}
\author{Zhu-an Xu}
\affiliation{School of Physics, Zhejiang University, Hangzhou 310027, China}
\author{Wei Li}
\affiliation{Institute of Theoretical Physics, Chinese Academy of Sciences, Beijing 100190, China}
\author{Jinkui Zhao}
\email{jkzhao@gbu.edu.cn}
\affiliation{School of Physical Sciences, Great Bay University, Dongguan 523808, China}

\author{Hanjie Guo}
\email{hjguo@sslab.org.cn}
\affiliation{Songshan Lake Materials Laboratory, Dongguan 523808, China}



\date{\today}

\begin{abstract}
  Triangular-lattice systems host a variety of ground states, ranging from quantum spin liquids to magnetically ordered phases, the latter of which can exhibit a sequence of magnetic phase transitions under applied magnetic fields. Here, we report magnetic and thermodynamic measurements, combined with powder and single-crystal neutron diffraction, on a high-spin, nearly isotropic Mn$^{2+}$ triangular-lattice system \KMSO. The compound undergoes long-range magnetic ordering below $T_\mathrm{N} \sim 4$~K in zero field. Contrary to expectations for an ideal Heisenberg system, the compound adopts an up-down-zero (UD0) magnetic structure down to the lowest temperature (0.05 K), rather than the commonly expected Y-type structure. This UD0 state is, however, highly sensitive to external magnetic fields. For fields applied along the $c$ axis, it is readily destabilized and replaced by the Y-type structure, followed by an up-up-down (UUD) phase corresponding to the 1/3 magnetization plateau. In contrast, when the field is applied within the triangular plane, the system evolves into a canted Y state at a higher critical field. These results reveal that weak anisotropy, though small in magnitude, exerts a strongly orientation-dependent influence, playing a key role in selecting the field-induced phases in this frustrated magnet.
\end{abstract}


\maketitle

\section{Introduction}

Two-dimensional triangular-lattice antiferromagnets are prototypical frustrated systems that can host a variety of intriguing phases, such as spin liquids and spin supersolids. In a seminal work, Wannier showed that Ising spins on a triangular lattice form a highly degenerate ground state with a residual entropy of 0.3383\textit{R} \cite{Wannier1950}. Anderson, on the other hand, suggested that the Heisenberg spins would pair to form a resonant valence state (RVB), resisting ordering even at zero kelvin \cite{Anderson1973,Anderson1987}. Although later studies favored a 120$^\circ$ ordered state both classically and quantum mechanically for the nearest-neighbor Heisenberg model \cite{Capriotti1999}, inclusion of next-nearest-neighbor interactions \cite{Iqbal2016} or anisotropy \cite{Zhu2018} may still give rise to a RVB or quantum spin liquid state. On the ordered side, depending on the anisotropy and the application of magnetic field along different directions, the triangular-lattice system can show rich phases such as the so-called Y state, the up-up-down (UUD) state, and the V state \cite{Kawamura1985,Chubukov1991,Yamamoto2014,Yamamoto2019}. Notably, in a uniaxial system, the \textit{z} and \textit{xy} spin components of the Y and V states break the translation and U(1) rotation symmetries, respectively, representing a magnetic analog of the supersolid state \cite{Boninsegni2012}. However, experimental realization of the easy-axis anisotropy in the triangular-lattice transition-metal systems remains scarce until the discovery of Na$_2$BaCo(PO$_4$)$_2$ \cite{Zhong2019,Sheng2022,Gao2022,Xiang2024,Sheng2025} and K$_2$Co(SeO$_3$)$_2$ \cite{Zhu2024,Chen2026} systems recently. At low temperatures, these systems exhibit a 1/3 magnetization plateau in modest magnetic fields, where the excitation spectra obtained from neutron scattering measurements can be well described by an XXZ model with easy-axis anisotropy. Upon decreasing the field, however, the sharp magnon dispersions evolve into continua, the origin of which remains under active  debate, with possible explanations including two-magnon excitation and proximity to a quantum spin liquid phase \cite{Sheng2022,Zhu2024,Sheng2025,Chen2026}. The zero-field magnetic ground state for these systems is suggested to be the Y state, thus, forming a spin supersolid phase. However, we note that there is no unambiguous determination of the magnetic structure in Na$_2$BaCo(PO$_4$)$_2$ and K$_2$Co(SeO$_3$)$_2$. In Na$_2$BaCo(PO$_4$)$_2$, up to five different magnetic structures, including the Y structure, can account for the observed experimental data due to the limited number of reflections \cite{Sheng2022}. Similarly, no definitive refinement of the magnetic structure of K$_2$Co(SeO$_3$)$_2$ has been reported to date.

In such a context, studies on relevant systems may provide new insights into the nature of the ground state and unusual excitation spectra. Neutron diffraction measurements on a high-spin analog of Na$_2$BaCo(PO$_4$)$_2$, i.e., Na$_2$BaMn(PO$_4$)$_2$ single crystal reveal complex spin structures under magnetic fields \cite{Biniskos2025}, and the importance of the interlayer interaction in stabilizing the different phases. A recent report of the magnetic excitations of K$_2$Mn(SeO$_3$)$_2$ showed a similar excitation continuum coexisting with single magnon excitations, which is attributed to magnon-magnon interactions for the high-spin system \cite{Zhu2026}. Also, in this work, a non-collinear Y state is proposed for the zero-field ground state.

In this paper, we perform comprehensive magnetic and thermodynamic, together with powder and single-crystal neutron diffraction measurements on \KMSO, demonstrating that the zero-field ground state is an up-down-zero (UD0) spin structure where one third of the moments remain disordered down to the lowest temperature.
Such a partially disordered state is unexpected for a nearly Heisenberg triangular-lattice system and suggests a subtle interplay of competing interactions. We argue that it is stabilized by a weak single-ion anisotropy together with interlayer couplings that promote three-dimensional ordering, as indicated by the critical behavior of the phase boundary.
This UD0 phase is, however, highly sensitive to external fields. A small magnetic field applied along the \textit{c} axis rapidly overcomes the anisotropy and drives the system toward Heisenberg-like behavior. In contrast, for in-plane fields, the system evolves into a canted Y phase rather than the inverted-Y state expected for an ideal Heisenberg model. These results highlight that even weak anisotropy, though seemingly negligible, can play a decisive role in selecting magnetic ground states and governing their evolution in frustrated systems.

\section{Experiment}
\begin{figure}
\centering
   \includegraphics[width=0.95\columnwidth]{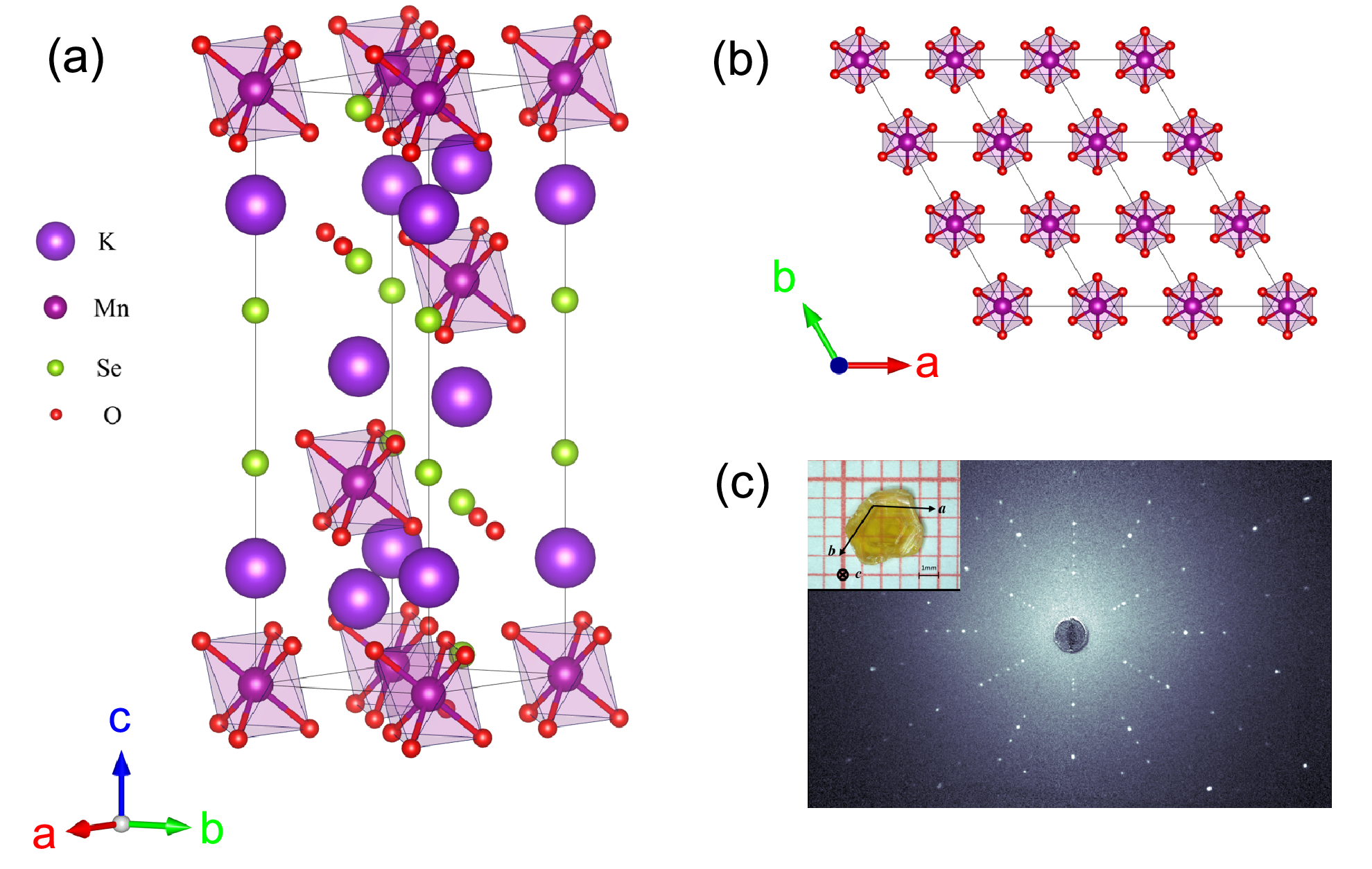}
   \caption{(a) Crystal structure of K$_{2}$Mn(SeO$_{3}$)$_{2}$. (b) Triangular network of the MnO$_6$ within the \textit{ab} plane. (c) Laue pattern of the single crystal with X-ray beam along the \textit{c} axis. The inset shows a photograph of the single crystal.}
   \label{fig1}
\end{figure}

Single crystals of K$_{2}$Mn(SeO$_{3}$)$_{2}$ were synthesized using the flux method. High-purity K$_{2}$CO$_{3}$ (99.99$\%$), MnO (99.5$\%$), and SeO$_{2}$ (99.99$\%$) were mixed and ground in a molar ratio of 1.2 : 0.95 : 2.2. The mixture was then loaded into a vacuum-sealed quartz tube and heated in a box furnace at a rate of 1~$^\circ$C/min to 600~$^\circ$C, and dwelled for 13 hours. Subsequently, the temperature was lowered at a rate of 0.5~$^\circ$C/min to 300~$^\circ$C, followed by natural cooling to room temperature. Finally, plate-like single crystals with hexagonal facets were obtained after dissolving the flux in distilled water. The typical crystal dimension is approximately 4 $\times$ 3 $\times$ 1 mm$^3$, as shown in the inset of Fig.~\ref{fig1}(c). The sharp spots in the Laue pattern indicate a high quality of the crystals; see Fig. \ref{fig1}(c).

Single-crystal X-ray diffraction (XRD) measurements were performed on a D8 VENTURE diffractometer (Bruker) equipped with a Mo X-ray source. The data reduction was performed using the APEX5 software.
The dc and ac magnetic susceptibility were measured using a Physical Property Measurement System (PPMS, Quantum Design). Heat capacity was measured on the same PPMS using the relaxation method.

Neutron powder diffraction measurements were performed on the High Intensity Powder Diffractometer (PKU-HIPD) at China Advanced Research Reactor, Beijing. About 2~g of pulverized \KMSO\ single crystals were loaded into a vanadium can and cooled down to 2.2~K. The incident neutron wavelength was 1.477 \AA.
Neutron single-crystal diffraction measurements were carried out on the SENJU time-of-flight (TOF) diffractometer in J-PARC, Japan. Two pieces of single crystals were aligned with their crystallographic \textit{c} and [1 -1 0] axes, respectively, along the vertical field (max. 6.8 T) direction. The neutron wavelength ranges from 0.7 to 4.4 \AA. The data reduction and integration were performed using the in-house program of the beamline. The (magnetic) structure refinements were performed using the \textit{Jana2020} \cite{Jana} and \textit{FullProf} suite \cite{Fullprof}. Additional magnetic symmetry analyses  were conducted using the Bilbao Crystallographic Server (BCS) \cite{bilbao1,bilbao2} and the ISODISTORT website \cite{isodistort}.

\section{Results}

\begin{figure*}
\centering
\includegraphics[width=1.95\columnwidth]{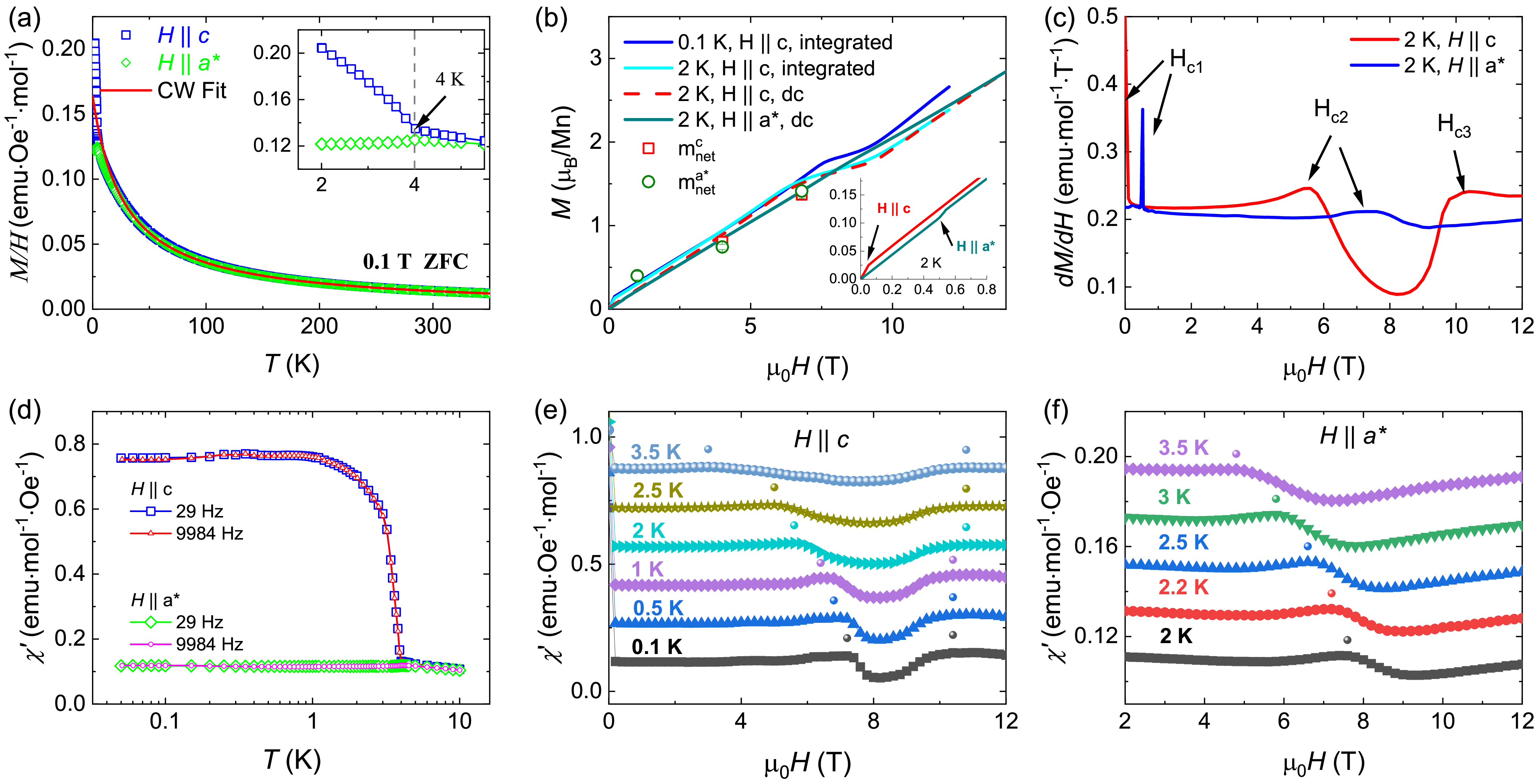}
\caption{(Color online) (a) Temperature dependence of the magnetic susceptibility measured with magnetic field applied along the \textit{c} and $a^*$ directions. The inset highlights the low-temperature region. (b) Isothermal magnetization for the fields applied along the \textit{c} and $a^*$ directions. The blue and cyan curves are obtained by integrating the ac susceptibility over magnetic field, as described in the text. The inset is an enlargement of the low-field region. (c) Magnetic field dependence of dM/dH. (d) Temperature dependence of the real component of the ac susceptibility, $\chi'$. (e,f) Magnetic field dependence of $\chi'$ measured at various temperatures with a driving frequency of 1429 Hz.  }
\label{sus}
\end{figure*}

Single-crystal X ray diffraction data reveal that K$_{2}$Mn(SeO$_{3}$)$_{2}$ crystallizes in the trigonal space group $R\bar{3}m$ (No.166), with lattice parameters $a=b=$ 5.5977(3)~\AA, \textit{c} = 18.5758(9)\AA, and angles $\alpha = \beta =$ 90$^{\circ}$, $\gamma = $ 120$^{\circ}$ at room temperature. The detailed experimental conditions and structural parameters can be found in the Supporting Information (SI). All atomic sites are fully occupied without indication of anti-site disorder or partial occupancy. The quality of the single crystals was further confirmed by powder X-ray and neutron diffraction measurements on crushed crystals, with no trace of any impurity phases.
The \KMSO\ compound consists of magnetic triangular networks of Mn$^{2+}$ ions, which are ABC-stacked along the crystallographic \textit{c} axis, as shown in Fig.~\ref{fig1}(a) and \ref{fig1}(b). The Mn$^{2+}$ ions are coordinated by six equivalent O atoms with a trigonal distortion, forming octahedra elongated along the \textit{c} axis. These MnO$_{6}$ octahedra are isolated from one another, sharing neither corners, edges, nor faces, and constitute the framework of the structure. Each Se$^{4+}$ ion is coordinated by three O atoms to form a typical SeO$_{3}$ trigonal pyramid, with Se-O bond lengths of about 1.689~\AA, consistent with those observed in common selenite (SeO$_{3}^{2-}$) compounds \cite{jo2018li6m,oh2012influence,k2004synthesis}. The SeO$_{3}$ units exhibit pronounced non-centrosymmetric character, likely due to the stereochemically active lone pair on Se$^{4+}$. The K$^{+}$ ions reside between the Mn-Se-O layers, serving both to balance charge and stabilize the structure.

Figure \ref{sus} summarizes the magnetic property measurements. Fig. \ref{sus}(a) presents the temperature dependence of the dc magnetic susceptibility, $\chi = M/H$, under a magnetic field of 0.1~T applied along the $a^*$ and $c$ directions. No bifurcation between the zero-field-cooled (ZFC) and field-cooled (FC) curves is observed. Thus, only the ZFC curves are shown. At high temperatures, the susceptibilities along the $c$ and $a^*$ directions overlap and increase gradually upon cooling. Below around 4~K, $\chi_c$ increases sharply, whereas $\chi_{a^*}$ decreases slightly, and then is almost temperature independent.  Curie-Weiss (CW) fits, $\chi(T) = C/(T-\theta_{CW})$, to the high-temperature data (150--350~K) yield $\mu_{\mathrm{eff}} = 6.06~\mu_B$ and $\theta_\mathrm{CW}$ = -28.6 K for $H \parallel a^*$, and $\mu_{\mathrm{eff}} = 6.09~\mu_B$ and $\theta_\mathrm{CW}$ = -28.3 K for $H \parallel c$. The effective moment is consistent with the expected value (5.92~$\mu_B$) for the high-spin $d^5$ configuration of Mn$^{2+}$ ($L = 0$ and $S$ = 5/2). The nearly identical CW temperatures indicate that the system is close to a Heisenberg limit with a nearest-neighbor antiferromagnetic exchange energy $J = 3\theta_\mathrm{CW}/nS(S+1)$ of 1.62 K (0.14 meV), where $n$ is the number of nearest-neighbor Mn$^{2+}$ ions within the plane. These results are consistent with a recent report \cite{Zhu2026}.

The isotherms are shown in Fig. \ref{sus}(b). When $H \parallel c$, a distinct magnetization plateau can be identified between $\sim$6 and 10 T at 2 K. The magnetization of the plateau amounts to about 1.64~$\mu_\mathrm{B}$/Mn, very close to 1/3 of the theoretical saturation moment (1.67~$\mu_\mathrm{B}$/Mn), indicating the presence of a UUD state. Such behavior is absent when $H \parallel a^*$, indicating the existence of anisotropy below $T_\mathrm{N}$. A closer inspection of the low-field data reveals additional features, with an abrupt increase of the magnetization around zero field when $H \parallel c$, and a kink at about 0.5 T when $H \parallel a^*$; see the inset of Fig. \ref{sus}(b). These meta magnetic transitions become more obvious in $dM/dH$ as shown in Fig. \ref{sus}(c). Although there is no obvious magnetization plateau in the MH curve when $H \parallel a^*$, a broad peak can be identified in $dM/dH$ at about 7.5 T and 2 K.

To probe further possible magnetic transitions below 2 K, we performed ac susceptibility measurements down to 50 mK. As shown in Fig. \ref{sus}(d), the real component of the ac susceptibility, $\chi'$, shows a sharp increase below $T_\mathrm{N}$, and becomes almost temperature independent below about 1 K when the field is along the \textit{c} direction.
Similar temperature-independent behavior is also observed when the field is applied along the $a^*$ direction, albeit with a magnitude of about 1/7 of that of $\chi'_c$, suggesting an easy-axis anisotropy along the \textit{c} axis. Moreover, it can be found that there is no frequency dependence along both directions, indicating that it is a long-range magnetically ordered state, rather than a spin glass state. Since there is no frequency dependence, we performed additional ac susceptibility measurements as a function of dc fields, from which the critical fields can be extracted as shown in Fig. \ref{sus}(e) and \ref{sus}(f). In addition, the integration of $\chi'$ over $H$ gives rise to the magnetization. As shown in Fig. \ref{sus}(b), the magnetization obtained in this way coincides with that from the dc measurements at 2 K. The 0.1~K data show that the plateau retains a magnetization of 1.84 $\mu_\mathrm{B}$/Mn, which is larger than that expected for 1/3 of the saturation moment, probably arising from the presence of van Vleck paramagnetism.

\begin{figure*}
\centering
\includegraphics[width=1.95\columnwidth]{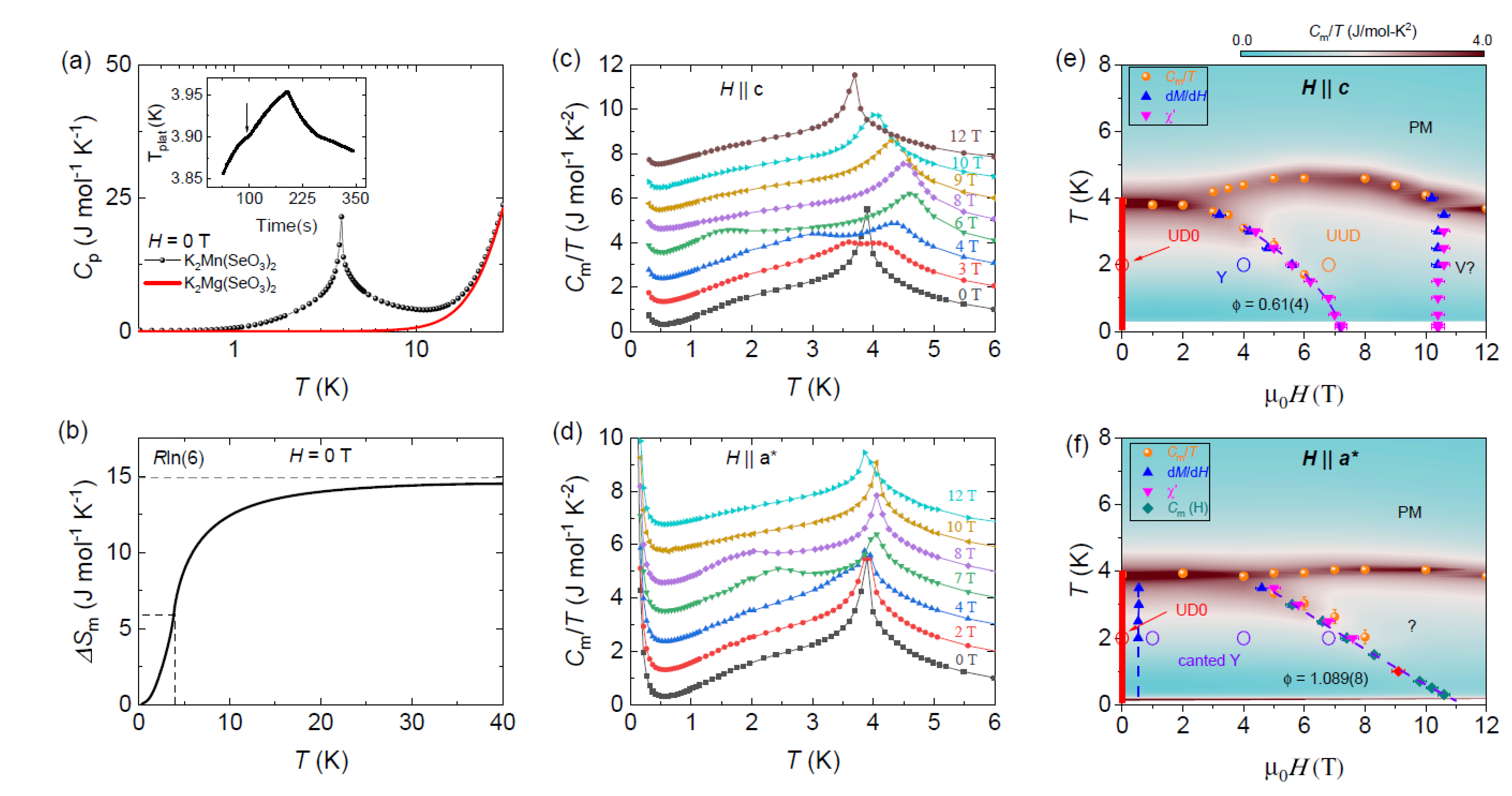}
\caption{(a) Temperature dependence of the zero-field specific heat for \KMSO. The data of K$_2$Mg(SeO$_3$)$_2$ is also shown as a phonon reference. The inset shows the temperature response curve near the first-order transition temperature during the specific heat measurement. (b) Magnetic entropy change obtained by integrating the magnetic specific heat $C_m/T$ over temperature. (c,d) Temperature dependence of the magnetic specific heat measured under different field strengths. (e,f) Magnetic phase diagram of \KMSO\ for field applied along the \textit{c} and $a^*$ directions, respectively. The circles indicate the positions where single-crystal neutron diffraction measurements were performed. The blue dashed line in (f) is an extrapolation based on the observation above 2 K. The violet dashed lines in (e) and (f) are critical fits as described in the text.}
\label{Cp}
\end{figure*}

To get more insights into the phase transitions, we performed specific heat measurements under different fields, as shown in Fig. \ref{Cp}. A sharp peak can be observed at $T_\mathrm{N}$ = 3.9 K, in line with the susceptibility measurements. A careful inspection of the temperature response curve around the transition temperature reveals a kink during the heating and cooling processes, suggesting a first-order transition with latent heat. The magnetic specific heat, $C_m$, is obtained by subtracting the phonon contributions using K$_2$Mg(SeO$_3$)$_2$ as a nonmagnetic analog. The entropy change, $\Delta S_m$, is obtained then by integrating $C_m/T$ over temperature and shown in Fig. \ref{Cp}(b). $\Delta S_m$ tends to saturate at 40 K to 14.5 J/(mol-K), close to the expected value for an $S$ = 5/2 state, indicating no further transitions are expected below 300 mK. Note that the entropy released above $T_\mathrm{N}$ amounts to 56\% of \textit{R}ln6, suggesting that short-range correlations develop well above $T_\mathrm{N}$.

When a field is applied along the \textit{c} axis, the peak at $T_\mathrm{N}$ splits into two peaks, with one moving toward higher temperatures and the other toward the opposite side. The low-temperature peak is more sensitive to the field and is almost indiscernible above 8 T. On the other hand, the high-temperature peak shifts to high temperatures first, and moves downward above 8 T, forming a dome shape in the phase diagram, as shown in Fig. \ref{Cp}(e).
Also note that the high-temperature peaks at 0 and 12 T are sharper than that in-between, indicating different nature for these transitions.
Similar splitting is also observed when the field is applied along the $a^*$ direction, albeit at higher critical fields. Moreover, the high-temperature peak remains sharp in this field direction. Combining the specific heat and magnetization measurements, a tentative phase diagram is drawn in Fig. \ref{Cp}(e) and \ref{Cp}(f) for fields along the \textit{c} and $a^*$ directions, respectively. Notably, the UUD phase region in Fig. \ref{Cp}(e) expands at higher temperature, suggesting that it is stabilized by thermal fluctuations. The persistence of a finite field range for the UUD phase down to near-zero temperature indicates that quantum fluctuations and/or magnetic anisotropy also play an important role \cite{Yamamoto2014}.

\begin{figure}
\centering
\includegraphics[width=0.85\columnwidth]{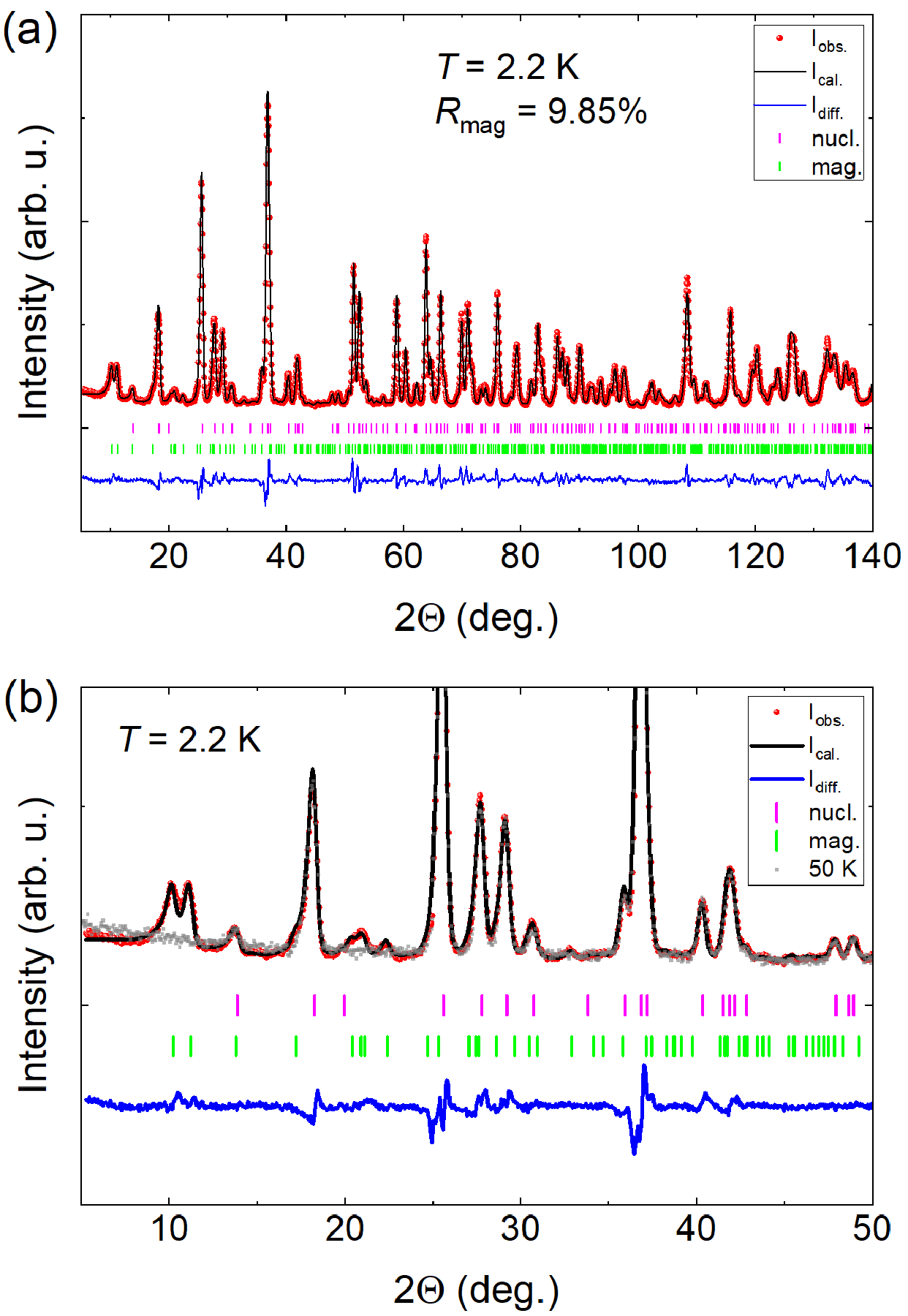}
\caption{(a) Simultaneous nuclear and magnetic structure refinements against the powder neutron diffraction pattern for \KMSO\ collected at 2.2 K. (b) Enlargement of the low-2$\theta$ region. The 50-K dataset is also shown for comparison.}
\label{NPD}
\end{figure}

To unveil the nature of these different phases, we performed powder and single-crystal neutron diffraction measurements at different fields.
Fig. \ref{NPD} shows the neutron powder diffraction pattern collected at 2.2~K. Compared to the 50-K data, additional reflections can be observed at 2.2~K, indicating a long-range magnetic ordering. These additional reflections can be indexed with a commensurate propagation vector \textbf{k}~=~(1/3~1/3~0); see Fig. \ref{NPD}(b). For magnetic symmetry analysis, we use the \textit{BasIreps} program within the \textit{FullProf} package. For the Mn ions at the 3\textit{a} site, the magnetic representation is decomposed into two irreducible representations (IRs) as:
$\Gamma = \Gamma_1 \oplus 2\Gamma_2$.
The basis vectors for these IRs are shown in Tab.~\ref{BV}. As can be seen, the moments are constrained along the [1~1~0] direction for $\Gamma_1$, while they are within the planes spanned by the [1~-1~0] and [0~0~1] directions for $\Gamma_2$.

\begin{table}
\caption{Irreducible representations together with basis vectors $\psi_\nu$ for Mn ions at the 3\textit{a} site with space group $R\bar{3}m$ and propagation vector \textbf{k}~=~(1/3 1/3 0). \label{BV}}
\begin{tabular}{lll}
IRs & $\psi_\nu$  & Mn \\
 \hline
$\Gamma_1$ & $\psi_1$  & (1~1~0)  \\
$\Gamma_2$ & $\psi_2$  & (1~-1~0) \\
           & $\psi_3$  & (0~0~1)  \\

\end{tabular}
\end{table}

The experimental data can be best described by $\Gamma_2$ with a combination of the two basis vectors. However, the propagation vector \textbf{k}~=~(1/3~1/3~0) can result in certain ambiguity in the magnetic structure. This can be best seen by considering that the moment at the \textit{l} lattice can be expressed as $\textbf{m}_l = \textbf{m}_0\mathrm{cos}(2\pi \textbf{k}\cdot\textbf{R}_l+\varphi_0)$, where $\textbf{R}_l$ is the lattice vector and $\varphi_0$ is an initial phase.
When $\varphi_0$ = 0, it will give rise to a magnetic structure of the up-down-down (U$\frac{D}{2}\frac{D}{2}$) type with the size of the down moment half that of the up moment.
On the other hand, if $\varphi_0$ = $\pi$/2, a partial disordered phase is realized, in which one third of the moments are disordered, while the remaining two thirds have equal magnitude but point in opposite directions -- dubbed as a UD0 phase. The former and latter structures would correspond to a magnetic space group $C2'/m'$ and $C2'/m$, respectively, as obtained from the BCS.
A refinement based on the U$\frac{D}{2}\frac{D}{2}$ model yields an up moment of 5.5(2) $\mu_\mathrm{B}$/Mn. This is larger than the maximum value for Mn$^{2+}$ (5 $\mu_\mathrm{B}$/Mn). On the other hand, the UD0 model gives rise to an equal size of moment for the U and D spins of 4.8(2)~$\mu_\mathrm{B}$/Mn. Therefore, based on the argument of moment size, the system is believed to order in the UD0 structure at zero field. Such a model differs from that expected for the triangular lattice Heisenberg model, which predicts a Y-type structure at zero field  \cite{Kawamura1985}. The Y phase can be constructed by a linear combination of the basis vectors of $\Gamma_2$ such that $\psi = A \psi_2 + iB \psi_3$ with $B/A = 1.73$. More generally, the spiral will have an elliptical envelope if $B/A \neq 1.73$. Such a model cannot describe the experimental data satisfactorily, especially for the (1/3~-2/3~1) peak at $2\Theta = 11.08^\circ$; see the SI for more details.

Single-crystal neutron diffraction measurements confirm the propagation vector \textbf{k} = (1/3 1/3 0) at zero field. When $H \parallel c$, the propagation vector remains unchanged up to 6.8 T; see Fig. S2 in the SI. On the other hand, when a field larger than 0.7 T is applied along the [1 -1 0] direction, an additional propagation vector \textbf{k}$^*$= (1/3~1/3~3/2) appears, indicating a modulation of the magnetic structure along the \textit{c} axis, and nonnegligible interactions between the triangular planes.

Refinement of the zero-field magnetic structure against the single-crystal data confirms the powder results, namely a UD0 structure rather than a U$\frac{D}{2}\frac{D}{2}$ or Y structure. Moreover, our refinements show the presence of three \textit{K} domains, which are connected to the three arms of the star [\textbf{k}$_1$ = (1/3 1/3 0), \textbf{k}$_2$ = (1/3 -2/3 0), and \textbf{k}$_3$ = (-2/3 1/3 0)]. To solve the magnetic structure in fields, we first consider the possible realization of a UUD phase at 6.8 T when the field is parallel to the \textit{c} axis.
Such a model would require the disordered moment at zero field to attain a nonzero moment. Moreover, our angular dependence of the magnetization in a field of 2~T indicates that the moment is nearly isotropic within the \textit{bc} plane at high fields; see Fig. S1 in the SI. Therefore, it could be expected that all the moments are along the \textit{c} axis, either parallel or antiparallel to the field direction.
Such a state would be consistent with the magnetic structure with magnetic space group $C2'$ based on the BCS analysis. This structure allows non-zero magnetic moments on the three sublattices, while remaining constrained within the [1 -1 0]-c plane. We impose further constraints that the moments are of the same size and parallel to the magnetic field direction (along the \textit{c} axis). The calculated intensities are found to be in good agreement with the observed ones. Moreover, the net moment is close to the magnetization measured at 2 K, as shown in Fig. \ref{sus}(b). In fact, the magnitude of the net moment imposes strong constraints on the magnetic structure. For example, such a collinear UUD structure can also describe the 4-T data. However, this would give rise to a net moment of 1.35 $\mu_\mathrm{B}$/Mn, much larger than the experimental observation. Thus, the two U moments should rotate away from the \textit{c} axis, resulting in the so-called Y structure. The refinement is satisfactory, and the net moment is in excellent agreement with the experiment; see Fig. \ref{sus}(b) and the SI for more details.

\begin{figure}
\centering
\includegraphics[width=0.9\columnwidth]{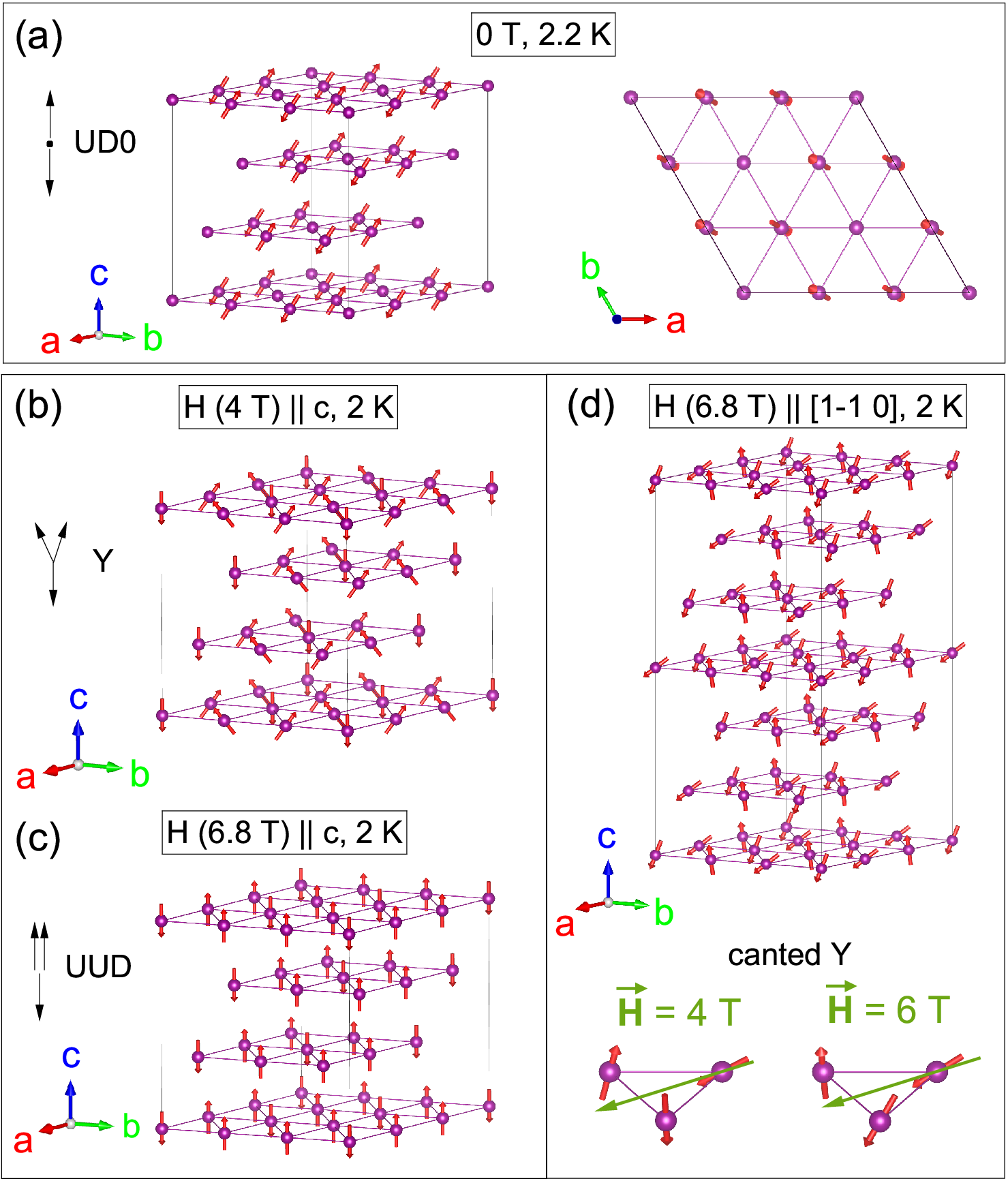}
\caption{Representative magnetic structures. (a) The UD0 structure at zero field. (b) The Y-type structure at 4 T when $H \parallel c$. (c) The UUD-type structure at 6.8 T when $H \parallel c$. (d) The canted Y-type structure at 6.8 T when $H \parallel$ [1 -1 0]. The lower insets show the rotation of the spins with increasing fields.}
\label{magstr1}
\end{figure}

Moving to the $H \parallel$ [1 -1 0] case. First we note that the applied magnetic field breaks the three-fold rotation, thus, no signature of domain formation is observed. Symmetry analysis by incorporating both the \textbf{k}$_1$ = (1/3~1/3~0) and \textbf{k}$^*$ = (1/3~1/3~3/2) propagation vectors results in sixteen maximal subgroups. Careful examination of these magnetic structures, excluding the ones with zero net moment etc. finally shows that only the magnetic space group $C2'/c'$ can describe the experimental data satisfactorily. The moments on the three sublattices are constrained to have the same size, but free to rotate within the [1 -1 0]-\textit{c} plane. The best refinements show that one of the three sublattice moments lies almost within the triangular plane, while the other two are nearly antiparallel and orient close to the \textit{c} axis. With increasing fields, these two moments progressively tilt toward the field direction. The representative magnetic structures can be found in Fig. \ref{magstr1}.

\section{Discussion}

A high-spin Mn$^{2+}$ system is expected to behave isotropically as a result of the quenched orbital moment to the first-order approximation. This seems also to be the case for \KMSO, as evidenced by the nearly identical Curie-Weiss temperatures along and perpendicular to the \textit{c} axis, and the weak angular dependence of the magnetization at 10 K. For such an isotropic high-spin system, classical Monte Carlo simulations suggest a 120$^\circ$ structure in zero field, followed by successive transitions to a UUD state, a V state, and finally a fully polarized state in fields \cite{Kawamura1985}. Similar phase transitions are also predicted for the Ising-like XXZ model both for the quantum and classical spins \cite{Yamamoto2014}. In addition, a UD0 or U$\frac{D}{2}$$\frac{D}{2}$ state can also be stabilized in the intermediate temperature range between the paramagnetic and Y states in an easy-axis system \cite{Miyashita1985,Miyashita1986,Melchy2009}. Such a UD0 state has also been observed in VCl$_2$ and VBr$_2$ \cite{Hirakawa1983}, and in EuCo$_2$Al$_9$ \cite{Xu2026} recently. It is interesting to note that all these systems have a relatively large spin number and the orbital moment is quenched. In such systems, the single-ion anisotropy is suggested to be more significant than the anisotropic exchange interactions \cite{Melchy2009}. In \KMSO, the elongated trigonal distortion of the MnO$_6$ octahedra is expected to give rise to an easy-axis single-ion anisotropy term $DS_z^2$ with a negative \textit{D} \cite{Hempel1976}, favoring moment alignment along the \textit{c} axis. However, our magnetic structure refinement shows that the moments are canted away from the \textit{c} axis by about 33$^\circ$, consistent with the angular-dependent magnetization measurements. Since the value of $D$ is expected to be very small (in the order of 0.001 meV for O$_6$ coordination) \cite{Duboc2009,Shu2023}, we propose that the interlayer interactions cannot be neglected; instead, they compete with the single-ion anisotropy and ultimately determine the moment orientation.

Evidence of the three-dimensional (3D) ordering can be drawn from the phase boundary analysis. In a uniaxial symmetry system, by mapping the long-range magnetically ordered state into a lattice of bosons, the UUD state corresponds to a 1/3-filled bosonic crystal, while the transverse (XY) ordering in the Y phase is associated with a Bose-Einstein condensation (BEC) \cite{Zapf2014}. The phase boundary should follow BEC universality such that $T_c \propto |H_c(0) - H_c(T)|^\phi$, where $\phi = z/d$ with dynamic exponent \textit{z} = 2, and $d$ the spatial dimensionality \cite{Zapf2014}. Analysis in the vicinity of the critical point when $H \parallel c$ yields $\phi$ = 0.61(4), and $H_c(0)$ = 7.18(5) T. The value of $\phi$ is close to the expected value of 2/3 for a three-dimensional (\textit{d} = 3) BEC. Similar analysis for $H \parallel a^*$ yields $\phi$ = 1.089(8). However, the application of a magnetic field along the $a^*$ direction breaks the uniaxial symmetry explicitly, and the magnetic structure above the critical field is unknown in the present study. Therefore, the nature of this nearly linear behavior when $H \parallel a^*$ remains unclear, which deserves further investigations in the future. The 3D character of the system may stabilize the UD0 phase down to the measured lowest temperature in zero field, as no additional apparent anomaly is observed in the specific heat and ac susceptibility. As a comparison, similar systems such as EuCo$_2$Al$_9$ \cite{Xu2026,Shu2026} and Na$_2$BaMn(PO$_4$)$_2$ \cite{Kim2022,Zhang2024,Biniskos2025} display a second anomaly below $T_\mathrm{N}$ in the specific heat, corresponding to the ordering of the XY component. Upon application of a magnetic field, however, the UD0 phase is rapidly suppressed, as evidenced by the abrupt changes of the isothermal magnetization curves at $H_{c1}$; see Fig. \ref{sus}(c). Note that the transition at $H_{c1}$ is much sharper than those at $H_{c2}$ and $H_{c3}$, indicating that the former is likely first-order, whereas the latter are second-order transitions. In addition, the critical field $H_{c1}$ along the \textit{c} axis is significantly smaller than that along the $a^*$ direction, indicating that despite the expected weakness of the anisotropy, it can still govern the field-induced behavior of the system.

\section{Conclusions}

In summary, we have performed a detailed investigation on the phase diagram of the nearly Heisenberg system \KMSO\ characterized by almost identical Curie-Weiss temperatures along and perpendicular to the \textit{c} axis. In zero field, the system adopts a UD0 magnetic structure down to the lowest temperature, rather than the Y-type structure expected for an ideal Heisenberg system. When a field is applied along the \textit{c} axis, the UD0 state is readily destabilized and gives way to the Y phase, followed by the UUD phase, and possibly the V phase at higher fields. When the field is applied within the triangular plane, the system evolves into a canted Y structure, also distinct from the inverted-Y structure expected for the ideal Heisenberg system. Moreover, we identify a three-dimensional BEC when the field is applied along the \textit{c} axis.
These results indicate that despite the weakness of the anisotropy in \KMSO\ with quenched orbital moment, the anisotropy plays a vital role in governing the magnetic behavior across the phase diagram. Our study highlights the rich variety of phases accessible in this classical spin system and suggests that real quantum magnets can exhibit unexpected behavior beyond idealized models.
It will also be interesting to explore how the remaining disordered moments in the UD0 phase will affect the spin dynamics. Specifically, will it give rise to excitation continua as observed in several related systems?
Future studies on Na$_2$BaCo(PO$_4$)$_2$ and K$_2$Co(SeO$_3$)$_2$, aiming at unambiguously determining their magnetic structures are highly desirable.

\section{Acknowledgement}
We thank Shang Gao for valuable discussions. This work was supported by the Guangdong Basic and Applied Basic Research Foundation (Grant No. 2022B1515120020, 2023B151520013), the National Research and Development Program of China (No. 2023YFF0718400, 2023YFA406501) and the Guangdong Provincial Quantum Science Strategic Initiative (Grant No. GDZX2501006). We gratefully acknowledge the neutron beamtime at the Materials and Life Science Experimental Facility of the J-PARC under a user program (Proposal No. 2025A0236).

\bibliography{KMnSeO}
\clearpage
\newpage

\renewcommand{\thefigure}{S\arabic{figure}}
\renewcommand{\thetable}{S\arabic{table}}
\setcounter{figure}{0}
\setcounter{table}{0}

\section*{Supporting Information for ``Directional selection of field-induced phases by weak anisotropy in triangular-lattice \KMSO"}

\section{Single-crystal XRD}

Single-crystal X-ray diffraction measurement conditions are listed in Tab. \ref{tab1}. The refined crystal structure parameters are tabulated in Tab. \ref{tab2}.

\begin{table}[h]
\centering
\caption{Experimental conditions for the single crystal XRD measurements, and agreement factors for the refinement.}\label{tab1}
\begin{tabular}{ll}
     \hline
			Formula & K$_{2}$Mn(SeO$_{3}$)$_{2}$ \\
			Space group & $R\ \bar{3}m$ (No. 166) \\
			$a, b$ (\AA) & 5.5977(3) \\
			$c$ (\AA) & 18.5758(9) \\
			$V$ (\AA$^3$) & 504.08(5) \\
			$Z$ & 2 \\
			$T$ (K) & 293 \\
			$\lambda$ (\AA) & 0.71073 \\
			$F$(000) & 537 \\
			$\theta$ (deg) & 3.29 - 28.19 \\
			$\mu$ (mm$^{-1}$) & 14.007 \\
            No. reflections  & 4335  \\
            No. independent reflections, $R_\mathrm{int}$  & 186, 6.03\%  \\
			No. of parameters & 16 \\
			Index ranges & $-7 \leq h \leq 7$, \\
			& $-7 \leq k \leq 7$, \\
			& $-24 \leq l \leq 24$ \\
			$R, wR_2$ [$I > 3\sigma(I)$] & 1.71\%, 4.13\% \\
			$R, wR_2$ [all data] & 2.05\%, 4.29\% \\
			Goodness of fit on $F^2$ & 1.3387 \\
			Largest difference peak/hole (e/\AA$^3$) & 0.53/-0.34 \\
\hline
\end{tabular}
\end{table}

\begin{table}[h]\caption{Refined crystal structural parameters for \KMSO.}\label{tab2}
\begin{tabularx}{\linewidth}{X X X X X X}
	\hline
	Atom  & $x$ & $y$ & $z$ & Occ. & $U_\mathrm{eq}$ (\AA$^{2}$)\\
	\hline
	K (6$c$)    & 0        & 1        & 0.19511(6)  & 1.00 & 0.0122(3) \\
	Mn (3$a$)    & 0        & 1       & 0           & 1.00 & 0.0051(3) \\
	Se (6$c$)    & 0.333333      & 0.666667        & 0.03858(2)  & 1.00 & 0.0057(2) \\
	O  (18$h$)   & 0.1749(2)  & 0.8251(2) & 0.07640(11) & 1.00 & 0.0099(7) \\
	\hline
\end{tabularx}

\begin{tabularx}{\linewidth}{X X X X X X X}
	\hline
	Atom & U11 (\AA$^{2}$) & U22 (\AA$^{2}$) & U33 (\AA$^{2}$) & U23 (\AA$^{2}$) & U13 (\AA$^{2}$) & U12 (\AA$^{2}$) \\
	\hline
	K (6$c$) & 0.0141(4) & 0.0141(4) & 0.0086(5)  & 0 & 0 & 0.0070(2) \\
	Mn (3$a$) & 0.0047(3) & 0.0047(3) & 0.0058(5)  & 0        & 0 & 0.00237(17) \\
	Se(6$c$) & 0.0054(2) & 0.0054(2) & 0.0063(3)        & 0  & 0 & 0.00271(12) \\
	O (18$h$)  & 0.0129(8) & 0.0128(8) & 0.0090(9) & 0.0002(4) & -0.0002(4) & 0.0103(9) \\
	\hline
	\end{tabularx}
\end{table}

\begin{figure}
\includegraphics[width=0.45\textwidth]{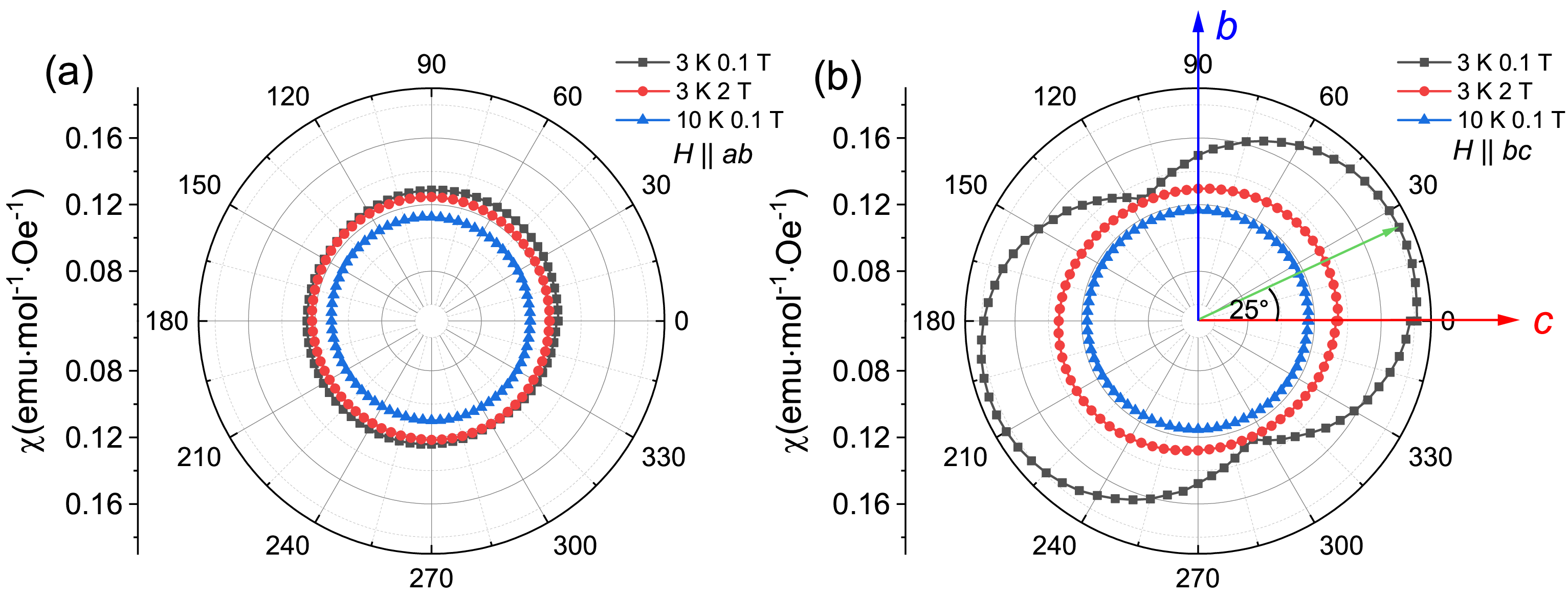}
\caption{Angular dependence of the magnetic susceptibility with magnetic fields applied (a) within the \textit{ab} plane, and (b) within the \textit{bc} plane.}
\label{angular}
\end{figure}

\section{Angular dependent magnetic susceptibility}

Angular dependence of the dc magnetic susceptibility is presented in Fig. \ref{angular}. As the field is rotated within the \textit{ab} plane, no apparent anisotropy could be observed, as shown in Fig. \ref{angular}(a). The anisotropy parameter, defined as $A = (\chi_\mathrm{max} - \chi_\mathrm{min})/(\chi_\mathrm{max} + \chi_\mathrm{min})$, amounts to 2.8\% at 3 K and 0.1 T, and 1.6\% at 10 K and 0.1 T. The small value of $A$ could arise from a slight deviation of the sample from the rotation center. In contrast, when the field is applied within the \textit{bc} plane, a large anisotropy is observed at low fields, reaching 17.6\% at 3 K and 0.1 T. However, upon increasing field, the anisotropy is significantly suppressed, with \textit{A} decreasing to 3.0\% at 3 K and 2 T, which is comparable to that for the field applied within the \textit{ab} plane. Also note that the maximum of $\chi$ at 0.1 T tilts about 25$^\circ$ away from the \textit{c} axis, consistent with our magnetic structure refinement at zero field.

\begin{figure}
\includegraphics[width=0.45\textwidth]{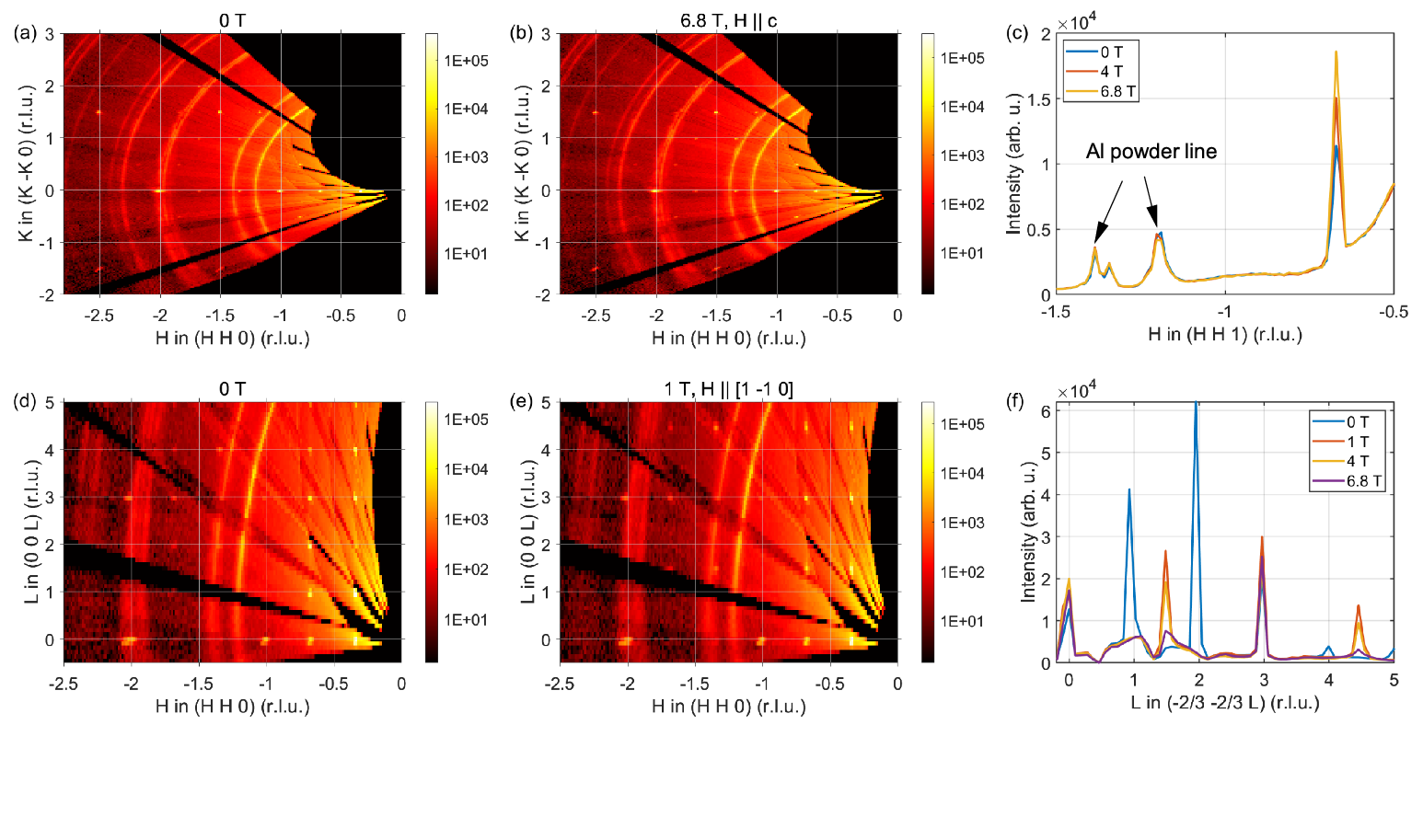}
\caption{Reciprocal spaces for magnetic fields applied (a,b) along the c axis, and (d,e) along the [1 -1 0] axis. (c) Intensity changes at different fields ($\parallel$ c) for a cut along the (H H 1) direction. (f) Intensity changes at different fields ($\parallel$ [1 -1 0]) for a cut along the (-2/3 -2/3 L) direction.}
\label{map}
\end{figure}

\section{Neutron diffraction}

\begin{figure}
\centering
\includegraphics[width=0.85\columnwidth]{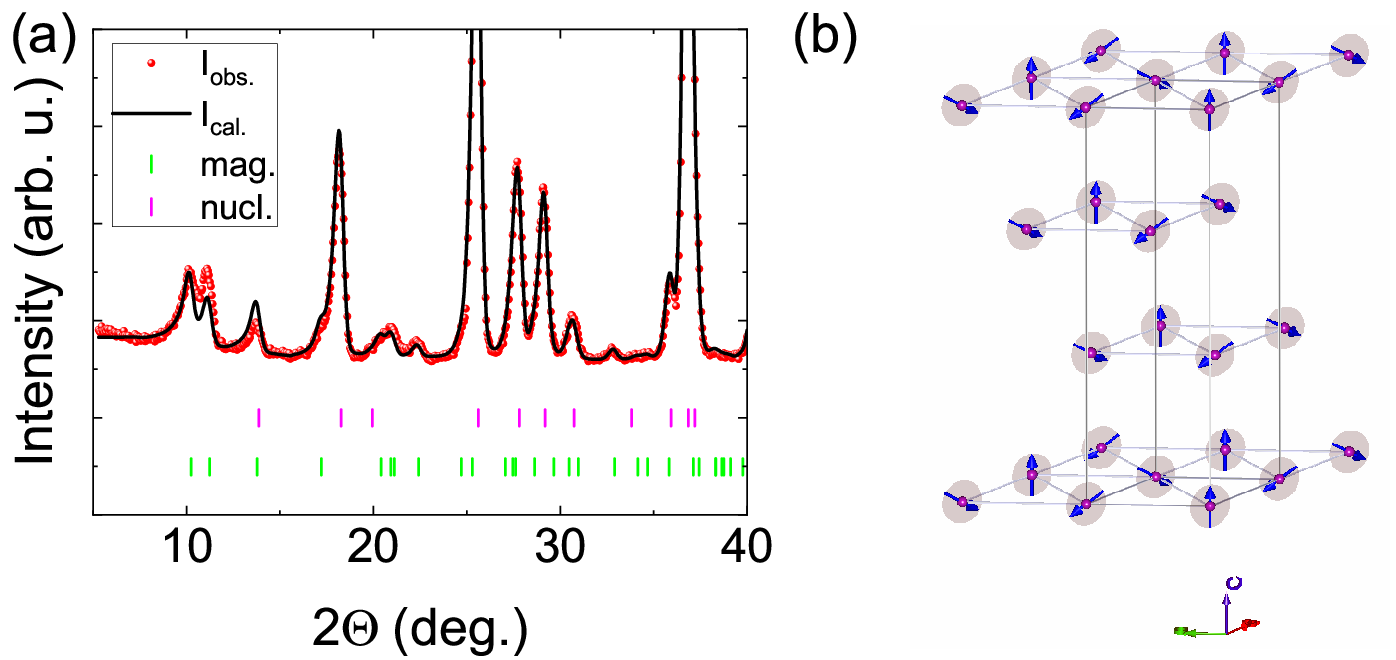}
\caption{(a) Refinement based on a Y-type structure. (b) The Y-type magnetic structure.}
\label{powder}
\end{figure}

\begin{figure}
\includegraphics[width=0.45\textwidth]{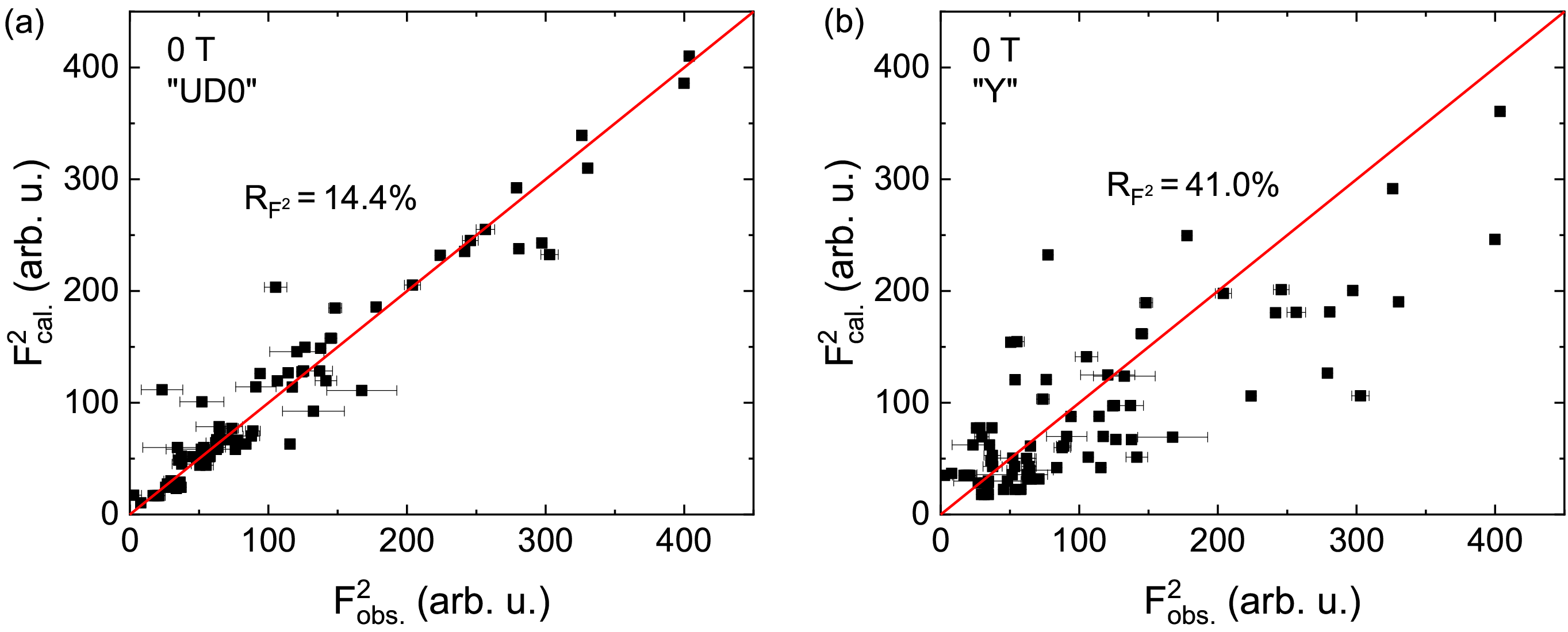}
\caption{Magnetic structure refinements based on (a) a UD0 and (b) a Y-type structure against the zero-field neutron single-crystal diffraction data.}
\label{single}
\end{figure}

\begin{figure}
\includegraphics[width=0.45\textwidth]{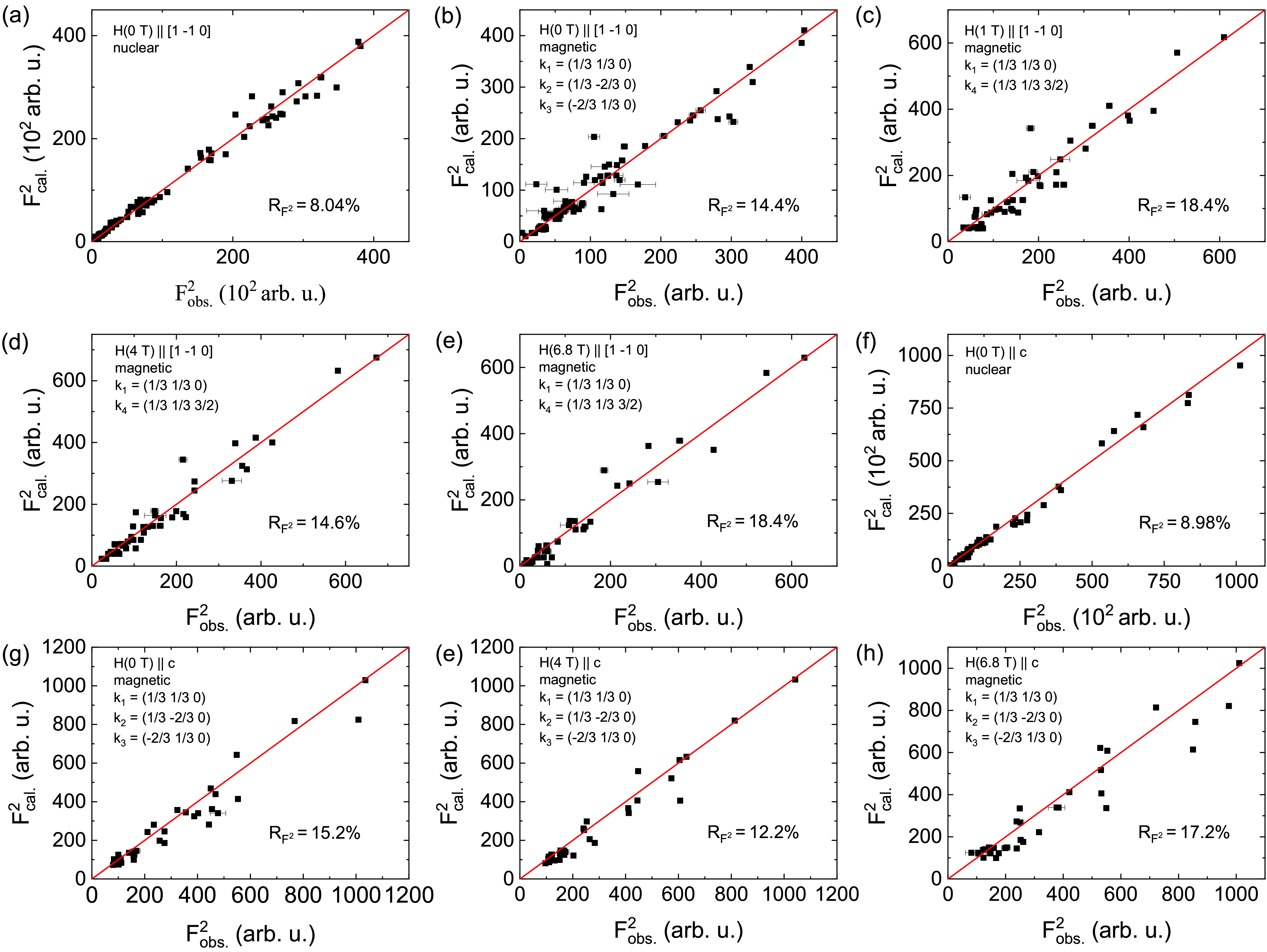}
\caption{$F_\mathrm{obs.}^2$ vs. $F^2_\mathrm{cal.}$ for the single-crystal refinements at different fields.}
\label{all}
\end{figure}

Single-crystal neutron diffraction measurements confirm that the propagation vector is \textbf{k}$_1$ = (1/3 1/3 0) and persists up to 6.8 T when a field is applied along the \textit{c} axis; see Fig. \ref{single}(a) and \ref{single}(b). On the other hand, a second propagation vector \textbf{k$^*$} = (1/3 1/3 3/2) appears when a field larger than 0.7 T is applied along the [1 -1 0] direction, indicating that the magnetic structure is modulated along the \textit{c} axis; see Fig. \ref{single}(d) and \ref{single}(e). As mentioned in the main text, the commonly expected Y state at zero field cannot describe the experimental data satisfactorily, which can be seen from the refinement against both the powder and single crystal data, as shown in Fig. \ref{powder} and \ref{single}. Refinements of the nuclear and magnetic structures against the single-crystal data are summarized in Fig. \ref{all}.

Moreover, it is found that there exists three \textit{K} domains which are related to the three arms of the star [\textbf{k}$_1$ = (1/3 1/3 0), \textbf{k}$_2$ = (1/3 -2/3 0), and \textbf{k}$_3$ = (-2/3 1/3 0)]. This can be easily seen from Tab. \ref{tab4} that one reflection corresponds uniquely to one arm.

\clearpage
\newpage

\begin{longtable}{rrr | l | l| l | l}
\caption{Part of the zero-field experimental intensities against the calculated values based on the structures corresponding to different arms of the \textbf{k}-star [\textbf{k}$_1$ = (1/3 1/3 0), \textbf{k}$_2$ = (1/3 -2/3 0), and \textbf{k}$_3$ = (-2/3 1/3 0)]. The indices are based on a magnetic supercell which is 3$\times$3$\times$1 larger than that of the nuclear unit cell.}\label{tab4}\\
\hline
h & k & l & F$^2_\text{obs}$ & F$^2_\text{cal}$ (k$_1$) & F$^2_\text{cal} ($k$_2$) & F$^2_\text{cal}$ (k$_3$) \\
\hline
\endfirsthead
\hline
h & k & l & F$^2_\text{obs}$ & F$^2_\text{cal}$ (k$_1$) & F$^2_\text{cal} ($k$_2$) & F$^2_\text{cal}$ (k$_3$) \\
\hline
\endhead

\endlastfoot

2   & -7 & 0   & 355.5266 & 345.3424 & 0.0000 & 0.0000 \\
-2  & -8 & 0   & 96.4208  & 0.0000   & 0.0000 & 101.6039 \\
-2  & -5 & 0   & 234.3764 & 0.0000   & 281.3656 & 0.0000 \\
1   & -8 & 0   & 210.3471 & 243.1102 & 0.0000 & 0.0000 \\
-2  & -8 & -1  & 161.2369 & 141.4094 & 0.0000 & 0.0000 \\
-2  & -8 & 0   & 83.2086  & 0.0000   & 0.0000 & 101.6133 \\
-1  & -7 & -1  & 274.7253 & 245.9569 & 0.0000 & 0.0000 \\
-1  & -7 & 0   & 256.1853 & 0.0000   & 198.0493 & 0.0000 \\
-7  & -4 & -1  & 99.9465  & 125.7241 & 0.0000 & 0.0000 \\
-7  & -4 & 0   & 80.0350  & 0.0000   & 72.8453 & 0.0000 \\
-7  & -4 & 1   & 97.4491  & 0.0000   & 0.0000 & 90.6923 \\
-4  & -7 & -1  & 101.0632 & 0.0000   & 74.8448 & 0.0000 \\
-4  & -7 & 0   & 96.9614  & 0.0000   & 0.0000 & 88.2686 \\
-2  & -5 & -1  & 387.3164 & 0.0000   & 0.0000 & 323.6774 \\
-2  & -5 & 0   & 442.7822 & 0.0000   & 281.6173 & 0.0000 \\
-2  & -5 & 1   & 552.8165 & 413.8557 & 0.0000 & 0.0000 \\
-8  & -2 & -2  & 168.0940 & 145.5193 & 0.0000 & 0.0000 \\
-8  & -2 & -1  & 141.2554 & 0.0000   & 0.0000 & 135.1183 \\
-8  & -2 & 0   & 110.4151 & 0.0000   & 83.8636 & 0.0000 \\
-8  & -2 & 1   & 157.9107 & 141.4244 & 0.0000 & 0.0000 \\
-8  & -2 & 2   & 107.4942 & 0.0000   & 0.0000 & 105.9314 \\
-4  & -1 & -1  & 548.1560 & 642.4387 & 0.0000 & 0.0000 \\
-4  & -1 & 0   & 323.5958 & 0.0000   & 356.9034 & 0.0000 \\
-4  & -1 & 1   & 449.6404 & 0.0000   & 0.0000 & 467.3441 \\
-5  & -5 & -1  & 158.7025 & 0.0000   & 99.6657 & 0.0000 \\
-5  & -5 & 0   & 274.6740 & 186.1052 & 0.0000 & 0.0000 \\
-5  & -5 & 1   & 157.7628 & 0.0000   & 0.0000 & 120.7677 \\
-4  & -4 & -1  & 262.5706 & 0.0000   & 0.0000 & 274.7761 \\
-4  & -4 & 0   & 402.0521 & 340.3418 & 0.0000 & 0.0000 \\
-4  & -4 & 1   & 296.9849 & 0.0000   & 226.7707 & 0.0000 \\
-2  & -2 & 0   & 767.2415 & 821.3732 & 0.0000 & 0.0000 \\
-4  & -4 & -1  & 327.5015 & 0.0000   & 0.0000 & 274.8106 \\
-4  & -4 & 0   & 477.3079 & 340.4058 & 0.0000 & 0.0000 \\
-4  & -4 & 1   & 358.2898 & 0.0000   & 226.7915 & 0.0000 \\
-2  & -2 & -1  & 454.5192 & 0.0000   & 361.0761 & 0.0000 \\
-2  & -2 & 0   & 1008.9041 & 824.3594 & 0.0000 & 0.0000 \\
-2  & -2 & 1   & 468.0671 & 0.0000   & 0.0000 & 437.5054 \\
-1  & -1 & 0   & 1035.1201 & 1038.7485 & 0.0000 & 0.0000 \\
\hline
\end{longtable}

\end{document}